\documentclass[twocolumn,12pt,tightenlines,superscriptaddress,notitlepage]{revtex4-1}

\usepackage{amsmath}
\usepackage{amssymb}
\usepackage{latexsym}

\usepackage{url}
\usepackage{xcolor}

\usepackage{graphicx}
\definecolor{newcolor}{rgb}{.8,.349,.1}

\usepackage{multirow}

\begin{document}



\begin{abstract}


For sufficiently short relativistic-intensity laser pulses, the disparity in time scales for electron and ion motion causes ions to behave as a fixed, neutralizing background. As the pulse duration or intensity is increased, ion motion becomes important, leading to instability in uniform plasmas but more complex, and potentially desirable behavior in structured targets. In this work, we introduce a new self-organized regime in laser-driven microchannels wherein ion motion facilitates stronger peak fields and high charge and photon conversion efficiency. 3-D particle-in-cell simulations demonstrate that the qualitative laser-microchannel interaction regime is governed by similarity parameters relating the pulse duration, spot size, and intensity to channel scales. The observed similarity suggests that lower-intensity experiments can inform designs for next-generation facilities, where the high-field, highly-radiative properties of the self-organized regime are especially desirable.

\end{abstract}


\title{Ion-motion-driven enhancement of energy coupling and stability in relativistic laser–microchannel interaction}

\author{K. Weichman}
\email[corresponding author, ]{kweic@lle.rochester.edu}
\affiliation{University of Rochester, Laboratory for Laser Energetics, Rochester, NY 14623, USA}
\author{M. VanDusen-Gross}
\affiliation{University of Rochester, Laboratory for Laser Energetics, Rochester, NY 14623, USA}
\author{G. Bruhaug}
\affiliation{University of Rochester, Laboratory for Laser Energetics, Rochester, NY 14623, USA}
\affiliation{Los Alamos National Laboratory, Los Alamos, NM 87545, USA}
\author{J. P. Palastro}
\affiliation{University of Rochester, Laboratory for Laser Energetics, Rochester, NY 14623, USA}
\author{M. Wei}
\affiliation{University of Rochester, Laboratory for Laser Energetics, Rochester, NY 14623, USA}
\author{A. Haid}
\affiliation{General Atomics, San Diego, California 92816, USA}
\author{A. V. Arefiev}
\affiliation{Department of Mechanical and Aerospace Engineering, University of California at San Diego, La Jolla, CA 92093, USA}
\affiliation{Center for Energy Research, University of California at San Diego, La Jolla, CA 92037, USA}
\author{H. G. Rinderknecht}
\affiliation{University of Rochester, Laboratory for Laser Energetics, Rochester, NY 14623, USA}

\maketitle

\section*{Introduction} 


High-intensity laser interaction with structured targets has gained significant interest in recent years due to the beneficial effects of target structuring on the energetic particles, radiation, and strong fields produced by the interaction~\cite{klimo2011nanosphere,cerchez2013grating,kluge2012conefoil,jiang2016nanowire,ji2016tube,stark2016gamma,curtis2018nanofusion,snyder2019structured,wang2019faraday, rinderknecht2021microchannel,bailly2020tubeion}
as well as advances in target manufacturing capabilities such as 3-D printing~\cite{Kawata2001-2pp,zhou2015-2pp}. 
Following ionization by an intense laser pulse, structured targets can comprise regions of both relativistically underdense and overdense plasma, as defined relative to the relativistically adjusted critical density $\gamma n_c \sim a n_c$, where $n_c = m_e \omega^2/4\pi e^2$ and $a = |e|E/mc\omega$ is the local magnitude of the envelope of the normalized vector potential for a pulse with frequency $\omega$.
In such targets, the spatially-varying density profile influences not only the propagation of the laser pulse but also the transfer of its energy into other forms.
For example, the relativistically overdense regions in a target with few-wavelength-scale structure may serve as sources of electrons or ions to be accelerated within the underdense regions \cite{kluge2012conefoil,ji2016tube,klimo2011nanosphere,wang2019ionmotion,chintalwad2025cone}. The target structure also shapes the overall electromagnetic field responsible for charged particle acceleration and radiation generation by guiding, focusing, or reflecting the incident laser pulse and through the generation of strong transitory or quasi-static fields \cite{xiao2016tubedla,ji2016tube,jiang2016nanowire,stark2016gamma,wang2019faraday,chintalwad2025cone}. Consequently, target structuring has the potential to facilitate the development of platforms for studying quantum electrodynamics (QED) \cite{he2021linearbw,he2022channelpairdouble} and laboratory astrophysics \cite{chen2023labastro}.

However, obtaining the benefits of target structuring depends on experimentally accessing a favorable regime, not only in the target's initial \textit{spatial} structure, but also on its \textit{temporal} evolution.
Therefore, regimes in laser--structured-target interaction depend not only on how the spatial scales of the laser pulse (e.g., wavelength and spot size) compare to the initial target parameters, but also on how the pulse duration compares to the time scale over which the target structure evolves.


\begin{figure*}
    \centering
    \includegraphics[width=0.95\linewidth]{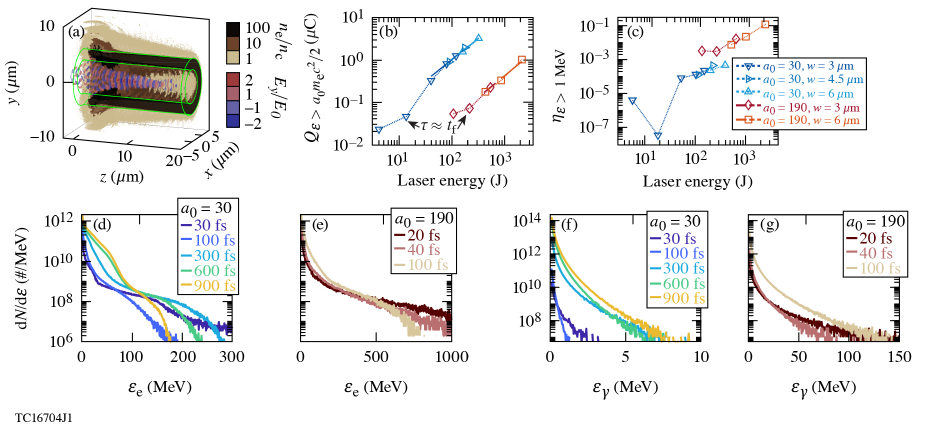}
    \caption{ A relativistic laser pulse interacting with an initially hollow microchannel produces copious energetic electrons and photons. (a) Snapshot of the high electron density and robust laser focusing ($E_y > E_0$, where $E_0$ is the peak field at focus in vacuum) that results from long pulses and mobile ions, from the highest-energy simulation in (b).
    (b),(c) Scaling of the total high-energy ($\varepsilon_e > a_0 m_e c^2/2$) charge (b) and the total energy conversion efficiency into energetic ($\varepsilon_\gamma>1\,\mathrm{MeV}$) photons (c) versus laser energy. The laser energy is varied by changing the pulse duration at fixed amplitude and spot size ($w$, FWHM of intensity).
    The solid lines in (b) correspond to Eq.~\ref{eqn:ntot-long}. 
    (bottom row) Energy spectra for (d),(e) electrons and (f),(g) photons for $w= 3\,\mu$m with (d),(f) $a_0 = 30$ and (e),(g) $a_0 = 190$.
    }
    \label{fig:summary}
\end{figure*}

Consider, for example, the interaction of a relativistic (peak vector potential $a_0 > 1$) laser pulse with an initially hollow microchannel (illustrated in Fig.~\ref{fig:summary}a). Such preformed microchannel targets, which consist of an overdense cylindrical tube filled with vacuum, gas, or foam have been proposed as a next-generation source of high-energy electrons, photons, ions, and positrons and as a platform for strong magnetic field generation \cite{kluge2012conefoil,xiao2016tube,xiao2016tubedla,stark2016gamma,jansen2018pair,snyder2019structured,wang2019faraday,bailly2020tubeion,rinderknecht2021microchannel,he2021channelpairdouble}. 
The dynamics of charged particle acceleration and QED processes depend on the field configuration within the channel~\cite{kluge2012conefoil,xiao2016tubedla,gong2019dla,gong2020channel,valenta2024parabolic}, and are thereby sensitive to evolution of the channel structure. 

Notably, the onset of channel evolution during the laser pulse duration, corresponding to the inward streaming of ions from the channel walls, has been observed to upend the initial electron acceleration and photon generation processes \cite{wang2019ionmotion,wang2020ioncomment}. The regime that emerges when the pulse duration is comparable to the time scale for ion motion has poor electron and photon energy, conversion efficiency, and angular divergence \cite{wang2020ioncomment}. However, avoiding ion motion altogether would confine experiments to short pulses or high-Z materials, and would be especially constraining at ultra-high laser intensity ($a_0 \gtrsim 100$).

In this paper, we demonstrate the emergence of a new, self-organized regime of laser-mircochannel interaction resulting from ion motion. In this long-pulse regime, the rising edge of the laser pulse pre-fills the channel with plasma, leading to stable self-focusing. The laser spot size controls the strength of the self-focusing and can be used to select between a rapid, efficient electron acceleration and photon generation process, and a more gradual process with lower angular divergences. 
The self-organized regime occurs for both high intensity ($1 < a_0 < 100$) and ultra-high intensity ($a_0 > 100$) and is characterized by similarity parameters related to the spatial and temporal scales of ions in the microchannel. 
The observed similarity suggests that experiments with structured targets using existing lower-intensity, longer-duration laser facilities ($\sim$100 J to 1 kJ in $\gtrsim 100$ fs)~\cite{maywar2008OMEGA_EP,Miyanaga2006LFEX,hopps2013orion,danson_vulcan_2004} can meaningfully inform future experiments at ultra-high intensity ($\sim$ kJ in $\lesssim 100$ fs)~\cite{weber2017eli-bl-10pw,tanaka2020eli-np-hpls,radier2022eli-np,nsfopalweb}, where the unique features of the long-pulse regime are particularly desirable.


\section*{Results and Discussion}

The ultimate goal of optimizing laser-microchannel (or laser-structured-target) interaction is to transfer the energy of the laser pulse into other useful carriers, such as high-energy electrons and photons. 
To this end, considerable effort has been made to understand how the products of laser-microchannel interaction are affected by increasing the laser energy (e.g., through increasing $a_0$, the spot size, and the pulse duration) and the microchannel parameters that affect the energy coupling (e.g., the channel radius and initial channel fill) ~\cite{ji2016tube,stark2016gamma,yu2018microtube,snyder2019structured,gong2019dla,wang2020channelpower,rinderknecht2021microchannel,he2021channelpairdouble}. 
In particular, increasing the pulse duration or intensity into the regime where the channel structure begins to evolve during the interaction modifies the electron acceleration and photon generation processes~\cite{wang2019ionmotion,wang2020ioncomment,bailly2020tubeion}.
Past work has shown that this transition from short to intermediate time scales (relative to the time it takes for ions to move substantially into the channel) is detrimental~\cite{wang2019ionmotion,wang2020ioncomment,bailly2020tubeion}. 
If instead, longer time scales are achieved, through either longer pulse duration or higher intensity, a new self-organized regime emerges with high charge and high photon conversion efficiency, as shown to the right of the black arrows in Fig.~\ref{fig:summary}b.

The source of this improvement is the evolution of the overall channel structure. Over time, an initially hollow (or low-density-filled) channel will fill with plasma, driven by the inward expansion of ions from the channel wall. For sufficiently long pulses, this process ultimately establishes a quasi-steady state plasma profile with favorable properties for electron acceleration and photon production. The time scale for channel filling can be estimated from a simple physical picture, roughly following the approach in Ref.~\citenum{wang2019ionmotion}.

Consider an initially hollow microchannel of inner radius $R$ consisting of an overdense electron-ion plasma with ions of mass $M$ and charge state $Z$. When a laser pulse of amplitude $a_w$ (where the subscript $w$ indicates evaluation at the channel wall) interacts with the target, electrons are pulled out of the channel wall to compensate the laser field. As the electric field oscillates, electrons are alternately pulled out of and pushed into the wall. However, not all of the electrons extracted during one half of the laser cycle are returned during the second half, leading to the net motion of electrons into the channel. This net extraction of electrons continues until the sheath field produced by charge separation becomes comparable to the electric field of the laser pulse at the channel wall, $E_w = a_w m_e c \omega/|e|$. Approximating the channel fill as a uniform density $n_e$, filling therefore produces $n_e = \Gamma E_w/(2\pi |e| R)$, where $\Gamma$ is an order-unity numerical fitting factor.

Ions are then pulled into the channel by the charge-separation field $E_r = -2\pi |e| n_e r = -\Gamma E_w r/R$.
The instantaneous velocity of an ion can be found from $M \mathrm{d}v/\mathrm{d}t = Z|e|E_r$ and $\mathrm{d}r/\mathrm{d}t = v$,  yielding $v^2 = \Gamma Z|e|E_w (R^2-r^2)/(M R)$. The time it takes for ions to reach the center is then $t_f = (\pi/2) \sqrt{MR/(\Gamma Z|e|E_w)}$. Using the numerical fit parameter obtained in Ref.~\citenum{wang2019ionmotion} (equivalent to $\Gamma = 0.39$), the channel filling time is approximately
\begin{gather}
    t_f \equiv T \sqrt{\dfrac{\mu}{a_w} \dfrac{R}{\lambda}\dfrac{m_p}{m_e}},
\end{gather}
where $T$ is the laser period, $\lambda$ is the laser wavelength, and $\mu = M/(Zm_p)$ (with proton mass $m_p$).
In practical units, for $\lambda = 1 \,\mu\mathrm{m}$,
\begin{gather} \label{eqn:t_f}
    t_f = 143 \,\mathrm{fs} \, \sqrt{\dfrac{\mu R\left(\mu\mathrm{m}\right)}{a_w}}.
\end{gather}
A laser pulse is therefore considered short if the pulse duration $\tau \ll t_f$, intermediate if $\tau \sim t_f$, and long if $\tau \gg t_f$. 

The $t_f$ for protons given by Eq.~\ref{eqn:t_f} is in good agreement with the filling times observed in 3-D particle-in-cell simulations of CH channels. Simulations exploring the three duration regimes were performed using the open-source particle-in-cell code WarpX~\cite{fedeli2022warpx}, for both high-intensity ($a_0 = 30$) and ultra-high intensity ($a_0 = 190$). Details of the simulation configuration are given in the Methods section.

\subsection*{Short-pulse regime}

\begin{figure*}
    \centering
    \includegraphics[width=0.95\linewidth]{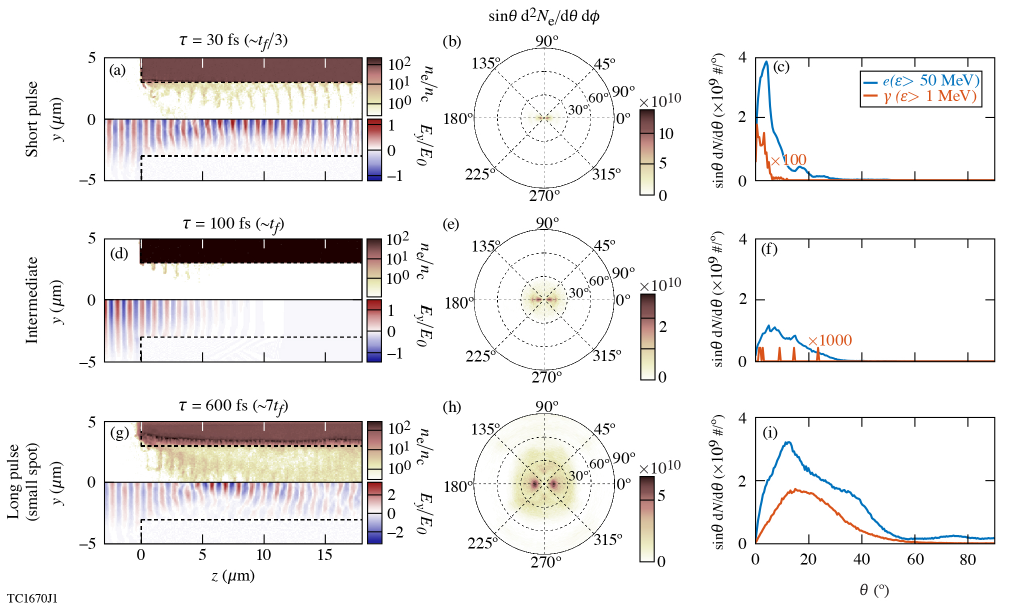}
    \caption{ 
    Ion motion leads to three pulse duration regimes, shown for $a_0 = 30$ and $w=3\,\mu$m. 
    (top row) Short pulse regime ($\tau < t_f$), characteristic of longitudinally-dominated DLA. (middle row) Intermediate regime ($\tau \sim t_f$) with inefficient electron acceleration and photon production. (bottom row) Long pulse regime ($\tau > t_f$) with robust electron acceleration and copious photon production due to channel pre-filling and laser focusing. (left column) Electron density and laser-pulse electric field strength when the peak of the pulse enters the channel. (middle column) Angular spectrum of electrons with energy above 50 MeV. (right column) Angular distribution of electrons and photons with respect to the forward direction. 
    $\theta = 0$ and $\phi=0$ are aligned with the $+z$ (laser propagation) and the $+y$ (laser polarization) directions, respectively.
    }
    \label{fig:regimes-a30}
\end{figure*}



In laser-driven microchannels, electrons gain energy via direct laser acceleration (DLA). The energy spectrum and divergence of both the accelerated electrons and the photons they radiate are sensitive to whether electrons gain energy predominantly from the longitudinal or the transverse field of the laser pulse. 
Longitudinal acceleration typically dominates in hollow plasma channels and is characterized by electrons accelerating in bunches along the channel walls, resulting in low-divergence electron and photon beams~\cite{xiao2016tubedla,gong2019dla}. Transverse DLA, on the other hand, typically dominates when the quasi-static self-generated fields are strong and produces channel-crossing electron trajectories, resulting in higher divergence~\cite{stark2016gamma,jansen2018pair,gong2019dla}. Strong quasi-static fields are generated in dense plasmas, causing transverse acceleration to dominate in microchannels with initial fill~\cite{kluge2012conefoil,xiao2016tubedla,gong2019dla,wang2020channelpower,gong2020channel}, in pre-formed parabolic plasma channels~\cite{vranic2018channel,babjak2024parabolic}, and in self-channeling by laser propagation in uniform (relativistically) near-critical plasma~\cite{pukhov_strong_2003,liu2013channelaccelcp,hu2015channelaccelcp,arefiev2016vphi,huang2017parabolic,hussein2021selfchanneldla}. In the short-pulse regime, the channel configuration -- and consequently the dominant acceleration pathway -- remains unchanged over the full duration of the laser pulse.

Figure~\ref{fig:regimes-a30}a shows a snapshot of electrons being accelerated in an initially hollow microchannel in the high-intensity short pulse regime, for a laser pulse with $a_0 = 30$ and $\tau = 30\,\mathrm{fs}$ ($t_f = 90\,\mathrm{fs}$, other parameters are given in Table~\ref{table:parameters}). Figs.~\ref{fig:regimes-a30}b and~\ref{fig:regimes-a30}c show the hallmarks of longitudinally-dominated DLA in the short-pulse regime: narrow-divergence electron and photon beams (Figs.~\ref{fig:regimes-a30}b and~\ref{fig:regimes-a30}c), with photons generated close to the channel wall. For the laser and channel parameters considered here, photon generation peaks in the sheath fields near the channel exit.

The effect of ion motion was studied by replacing the mobile CH with fixed ions. In the short pulse regime, ion mobility was found to have negligible effect on the electron energy and angular spectrum. However, modeling the ions as immobile significantly overestimates photon generation (by a factor of $\sim 3$ for the case shown in Fig.~\ref{fig:regimes-a30}c), due to the artificially increased strength of the sheath field.

\subsection*{Intermediate-pulse regime}

When the pulse duration is increased to $\tau \approx t_f$, significant motion of the channel wall begins to occur during the pulse duration. 
Filling of the channel changes the dominant electron acceleration process during the laser pulse from longitudinally-dominated to transversely-dominated~\cite{wang2019ionmotion} and can also pinch off the falling edge of the laser pulse. As shown in Figs.~\ref{fig:summary} and~\ref{fig:regimes-a30}, these effects are detrimental to both electron acceleration and photon generation.

\subsection*{Long-pulse regime} \label{sec:long-over}

Although ion-motion-driven evolution of the channel structure initially destabilizes electron acceleration and photon generation, a self-organized quasi-steady-state with improved performance emerges as the laser pulse duration is increased beyond the channel filling time.
As shown in Figs.~\ref{fig:summary}b and~\ref{fig:summary}c, this long-pulse regime has significantly higher total accelerated charge and photon conversion efficiency than the short-pulse regime.

The increase in the total accelerated charge is the result of the effective pre-filling of the channel by the laser-wall interaction. As shown in Fig.~\ref{fig:regimes-a30}g, near the channel entrance plane, the motion of ions causes inward expansion of the overdense channel wall. The number of energetic electrons pulled from the (collapsing) channel wall increases until the field due to charge separation becomes comparable to the electric field of the laser pulse. Considering now an approximately planar sheath of electrons with wavelength-scale thickness, (i.e., $|E_s| \sim 4\pi |e| n_e \lambda$), the density of electrons which interact with the laser can be approximated by $n_e \sim a_0 n_c/(2\pi)$. In this expression, $a_0$ appears rather than $a_w$ because the collapse of the wall pulls dense plasma into contact with the most-intense portion of the laser pulse. 

As this picture suggests, although the channel structure initially evolves on a temporal scale $t_f \ll \tau$, the subsequent, slow evolution of the laser field strength near the center of the channel (on the time scale of $\tau$) leads to the production of a self-organized, near-stationary structure that persists for the remainder of the pulse duration. Figure~\ref{fig:regimes-a30}g captures a snapshot of the quasi-equilibrium condition observed in simulations.

Using the plasma density scaling derived above, the total accelerated electron charge in the long-pulse regime is expected to scale as
\begin{equation} \label{eqn:ntot-long}
    Q = \xi |e| a_0 n_c c\tau w^2,
\end{equation}
where $\xi$ is introduced as a fit parameter.
The predictions of this scaling are compared to simulation results in Fig.~\ref{fig:summary}b for a variety of laser conditions. The fitting procedure is discussed in the Methods section. For the range of parameters considered in this work, good agreement is achieved with $\xi = 0.095$ for electrons with energy $\varepsilon_e/mc^2 > a_0/2$ and $a_0 < 100$. In the ultra-high-intensity regime, a larger fraction of the laser energy is ultimately transferred to ions and to photons, decreasing the $\xi$ observed after the interaction (e.g., $\xi \approx 0.025$ for the simulations conducted with $a_0 = 190$).

\subsection*{Dependence on spot size and channel radius} 

\begin{figure*}
    \centering
    \includegraphics[width=0.95\linewidth]{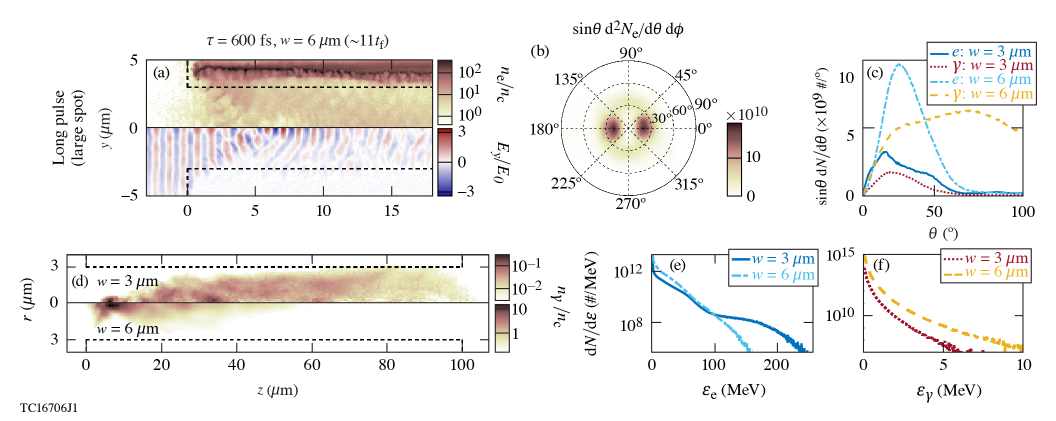}
    \caption{ The strength of interaction with the channel wall induces spot size dependence for long pulses, shown for $a_0 = 30$. (a),(b) Snapshot from the large-spot, long-pulse regime, with $w = 2 R$. 
    (a) Electron density and laser-pulse electric field strength when the peak of the pulse enters the channel. (b) Angular spectrum of electrons with energy above 50 MeV. (c),(d),(e),(f) Comparison between the large spot ($w=2R$) and smaller spot ($w=R$) cases. (c) Angular distribution of electrons and photons. (d) Photons binned by generation location. (e),(f) Energy distribution of (e) electrons and (f) photons.
    }
    \label{fig:spot-a30}
\end{figure*}

Although the total number of electrons extracted from the wall and accelerated to high energy depends on $a_0$ rather than $a_w$ in the long pulse regime, the processes for electron acceleration and photon generation (and thereby their energy and angular spectra) are sensitive to $a_w$. 
This sensitivity is evident when comparing simulations in which the spot size is comparable to the channel radius
(Figs.~\ref{fig:regimes-a30}g, ~\ref{fig:regimes-a30}h, and~\ref{fig:regimes-a30}i; $w=R=3\,\mu\mathrm{m}$) against those in which the spot size is larger (Figs.~\ref{fig:spot-a30}a, ~\ref{fig:spot-a30}b, and~\ref{fig:spot-a30}c; $w = 2R = 6\,\mu\mathrm{m}$).
In both cases, the interaction of the laser pulse with the channel wall leads to inward motion of the critical density surface, focusing the laser pulse and increasing the peak field amplitude (e.g., visible as $E_y/E_0>1$ in Fig.~\ref{fig:regimes-a30}g, which has $w=R$). 
However, the tightness of the focus depends on the strength of the interaction, leading to greater peak fields and pronounced pinching of the laser pulse as $a_w \to a_0$ (e.g., Fig.~\ref{fig:spot-a30}a, which has $w=2R$). 
This long-pulse, large-spot regime produces a pinched spot size on the order of the laser wavelength, enhancing the peak vector potential by $\sim 3$-$4\times$ for a channel with $R=3\,\mu\mathrm{m}=3\,\lambda$.

Consequently, both the electron acceleration and photon generation (Fig.~\ref{fig:spot-a30}d) transition from being distributed over the length of the channel to predominantly localized near the high-intensity pinch as $w/R$ increases. Whereas the distributed acceleration and emission processes at smaller $w/R$ are similar to those observed in initially filled channels in the short-pulse regime~\cite{stark2016gamma,wang2020channelpower}, the localized process for larger $w/R$ are more reminiscent of the gamma flash produced in the front-surface pre-plasma of overdense planar targets~\cite{nakamura2012flash,hadjisolomou2023flashreview}.

The pinching of the laser pulse leads to trade-offs between the longitudinally distributed acceleration and emission of the smaller-spot regime and the highly localized larger-spot regime.
In the larger-spot regime, the increase in the maximum field amplitude over that of the smaller-spot regime produces a higher number of moderate-energy electrons (Fig.~\ref{fig:spot-a30}e) and increases the generated photon energy and number (Fig.~\ref{fig:spot-a30}f). 
However, the strong pinching of the field also increases the angular divergence of both electrons and photons (Figs.~\ref{fig:spot-a30}c). 
Thus, the spot size and channel radius can be used to tune whether the electron acceleration and photon generation processes are high-intensity, high-yield, and highly localized or lower-divergence, lower-yield, and require a longer channel.

\subsection*{Similarity between high and ultrahigh-intensity and implications for future work}

\begin{figure*}
    \centering
    \includegraphics[width=0.95\linewidth]{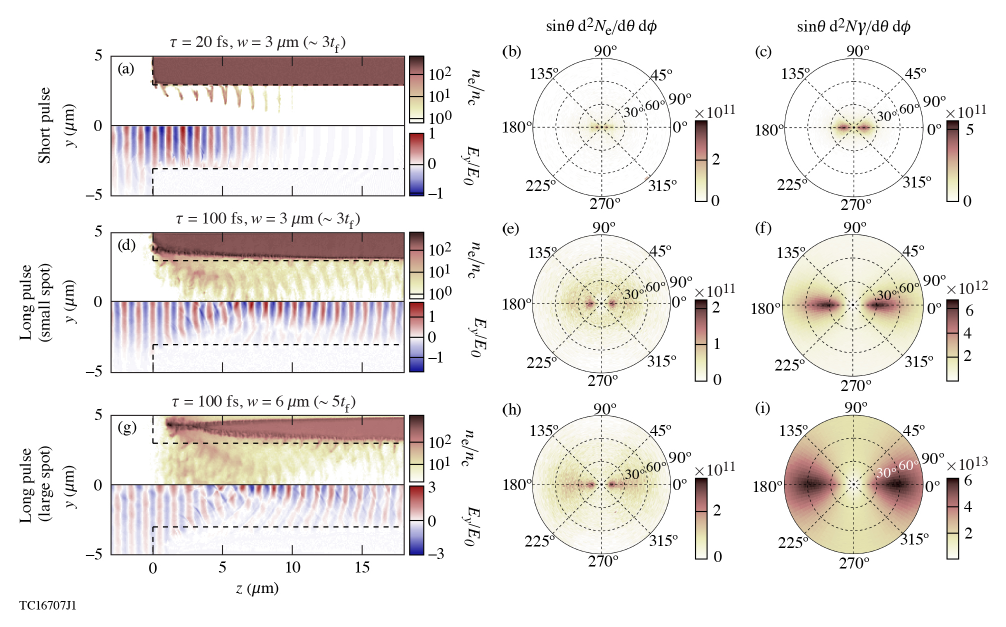}
    \caption{ Duration and spot size regimes for $a_0 = 190$. 
    (top row) Short pulse regime ($\tau < t_f$). (middle row) Long pulse regime ($\tau > t_f$) with $w=R$. (bottom row) Long pulse regime ($\tau > t_f$) with $w>R$. (left column) Electron density and laser-pulse electric field strength when the peak of the pulse enters the channel. (middle column) Angular spectrum of electrons with energy above 100 MeV. (right column) Angular spectrum of photons with energy above 1 MeV. 
    $\theta = 0$ and $\phi=0$ are aligned with the $+z$ (laser propagation) and the $+y$ (laser polarization) directions, respectively.
    }
    \label{fig:regimes-a190}
\end{figure*}

The duration and spot size regimes discussed in the preceding section are not unique to moderately relativistic pulses. As shown in Fig.~\ref{fig:regimes-a190}, simulations conducted with ultra-intense laser pulses (e.g., $a_0 = 190$) demonstrate the same ion-motion-mediated electron acceleration and photon generation regimes observed at lower intensity.
For example, Figs.~\ref{fig:summary}b and~\ref{fig:summary}c show the total accelerated charge and photon energy conversion efficiency are also improved in the long-pulse regime for ultra-intense pulses. 
As in the case of $a_0=30$, the angular divergence of electrons and photons for $a_0=190$ is increased when transitioning from the short- to the long-pulse regime (Figs.~\ref{fig:regimes-a190}b and~\ref{fig:regimes-a190}c vs Figs.~\ref{fig:regimes-a190}e and~\ref{fig:regimes-a190}f). 

Furthermore, the same tradeoffs as at lower intensity are observed when varying $w/R$. The smaller-spot regime similarly leads to longitudinally distributed electron acceleration and photon generation with lower-divergence but lower yield than the larger-spot regime (e.g., Figs.~\ref{fig:regimes-a190}d, \ref{fig:regimes-a190}e, and~\ref{fig:regimes-a190} with $w = R = 3\,\mu\mathrm{m}$ vs Figs.~\ref{fig:regimes-a190}g, \ref{fig:regimes-a190}h, and~\ref{fig:regimes-a190}i with $w = R = 6\,\mu\mathrm{m}$).
The latter case also produces similar enhancement of the peak field strength in both intensity regimes (e.g., $a\sim 3a_0$ in Fig.~\ref{fig:regimes-a190}g).

The similarity observed between the moderately-relativistic and ultra-relativistic regimes suggest that the qualitative features of the interaction are preserved as $a_0$ is varied with $\tau/t_f$ and $w/R$ held fixed. In this sense, $\tau/t_f$ and $w/R$ are similarity parameters governing the interaction of the laser pulse with the microchannel target. 
On the basis of this similarity, experiments conducted at lower intensity can be used inform the design of future ultra-high intensity experiments.
The similarity picture is expected to break down in the limit that QED becomes non-perturbative, due to the strong scaling of QED processes with $a_0$.

In conclusion, ion motion and the resulting evolution of the target configuration can have a beneficial impact on laser-structured target interaction. A relativistic-intensity laser pulse propagating in an initially hollow microchannel extracts electrons from the overdense channel wall, driving its inward expansion. This channel filling affects the electron acceleration and photon generation processes and leads to poor performance when the pulse duration is increased from short to intermediate relative to the channel filling time ~\cite{wang2019ionmotion,wang2020ioncomment}. However, ion motion self-organizes a quasi-stationary plasma structure that facilitates high charge and photon energy conversion efficiency for longer pulses. This structure focuses the laser pulse, increasing the peak field strength. The tightness of the focus is determined by the spot size of the laser pulse and the channel radius and controls whether the electron acceleration and photon generation processes are spatially localized (resulting in higher number) or longitudinally distributed (resulting in lower divergence). 

This work demonstrates that ion-motion-driven self-organized regimes in laser-structured target interaction can be both accessible and desirable for energetic particle sources and QED platforms. 
Achieving robust electron acceleration and high fields with both high-Z (nearly-stationary-ion) and low-Z (mobile-ion) structured targets could facilitate the study of Z-dependent QED processes such as Bethe-Heitler pair production and bremsstrahlung. In addition, self-organized pinching of laser pulses in structured targets could produce stable, extremely high fields for proposed ultra-high intensity systems such as NSF-OPAL~\cite{nsfopalweb}. Future work will evaluate the ability of microstructured targets to reach the non-perturbative QED regime and will seek to improve the photon angular divergence through increasing the efficiency of electron acceleration~\cite{tangtartharakul2025efficiency}.


\section*{Methods}

\subsection*{Simulations}

\begin{table}
\centering
\begin{ruledtabular}
\begin{tabular}{ l c }
  \multicolumn{2}{ l }{\textbf{Laser parameters} }\\
  Laser polarization & $y$ \\
  Propagation direction & $+z$ \\
  Wavelength & $\lambda=1$ $\mu$m \\
  Spot size (intensity FWHM) & $w=3\,\mu$m \\
  Duration (intensity FWHM) & $\tau = 30-900$~fs \\
  Focal plane & Channel entrance \\
  Peak amplitude & $a_0 = 30$ \\
  \hline
  \multicolumn{2}{ l }{\textbf{Channel parameters} } \\
  Inner radius & $R=3\,\mu$m\\
  Wall thickness & $2\,\mu$m \\
  Channel length & $100\,\mu$m \\
  Wall density & \multirow{2}{*}{$n_e = 100\,n_c$} \\
  (fully ionized CH) & \\
  \hline
  \multicolumn{2}{ l }{\textbf{Other parameters} }\\
  Simulation dimensionality & 3-D cartesian \\
  Spatial resolution & 30 cells/$\lambda$ \\
  Macroparticles per cell & \multirow{2}{*}{20/10/10}\\
  (electron/proton/carbon) &   \\ 
\end{tabular}
\end{ruledtabular}
  \caption{
  Nominal simulation parameters for $a_0=30$. The laser pulse is spatially and temporally Gaussian. For the simulations with $a_0=190$, the channel length was shortened to $40\,\mu$m, the wall density was increased to $200\,n_c$, and the number of particles per cell was doubled. 
  }
  \label{table:parameters}
\end{table}

Simulations were conducted using the open-source particle-in-cell code WarpX~\cite{fedeli2022warpx}. Nominal simulation parameters for the high intensity regime ($a_0 = 30$) are given in Table~\ref{table:parameters}. 
Additional simulations were performed for both high intensity and ultrahigh intensity ($a_0=190$) with decreased cell size and an increased number of particles per cell to ensure the resolution was adequate. The choice of computational algorithms (2nd-order finite difference time domain field solver with 2nd-order particle shape, Boris pusher with energy-conserving field gather and Galerkin interpolation, Esirkepov current deposition, and bilinear current filtering) ensured robust energy conservation and convergence at the employed resolution.

Quantum synchrotron photon emission was modeled using the Monte Carlo algorithm described in Ref.~\citenum{fedeli2022picsarqed}. Pair production was not included. The effect of pair production on the overall electron acceleration and photon generation processes was found to be negligible even in the highest-energy cases, via additional simulations including the nonlinear Breit-Wheeler pair production algorithm described in Ref~\citenum{fedeli2022picsarqed}. 

\subsection*{Fit parameter $\xi$ for total accelerated charge}

\begin{figure}
    \centering
    \includegraphics[width=0.95\linewidth]{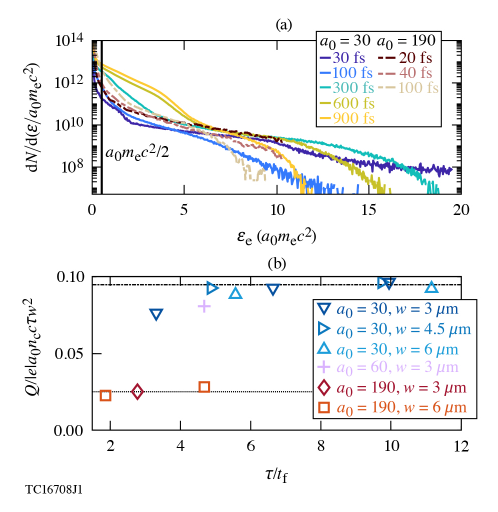}
    \caption{Charge produced by laser-microchannel interaction. (a) Energy spectra for different values of $\tau$ with $w=3\,\mu$m, showing energy cutoff shown for extraction of total charge (solid line). (b) Normalized total charge with energy above $a_0 m_ec^2/2$. The dashed and dotted lines represent Eq.~\ref{eqn:ntot-long}, with constant values of $\xi$ (0.095 and 0.025, respectively). 
    }
    \label{fig:fit}
\end{figure} 

The prediction of Eq.~\ref{eqn:ntot-long} was compared to the total charge with energy greater than $a_0 m_ec^2/2$ in simulations with varying $a_0$, spot size, and pulse duration. As shown in Fig.~\ref{fig:fit}a, the energy threshold $a_0 m_ec^2/2$ was chosen to capture the effect of increasing the pulse duration on the hot part of the electron energy spectrum while excluding the cold part of the spectrum. As shown in Fig.~\ref{fig:fit}b, the total charge observed in simulations is in good agreement with Eq.~\ref{eqn:ntot-long} for $\tau \gtrsim 2 t_f$, provided different values of $\xi$ are applied for $a_0=30$ and $a_0=190$. An additional simulation with $a_0 = 60$ suggests that the value of $\xi$ found for $a_0=30$ remains a good estimate for $a_0 \lesssim 100$. The lower value of $\xi$ in the ultra-intense regime (e.g., at $a_0 = 190$) is to be expected, as photons ultimately carry a significant fraction of the incident laser energy. 

\section*{Acknowledgments}

This material is based upon work supported by the Department of Energy [National Nuclear Security Administration] University of Rochester “National
Inertial Confinement Fusion Program” under Award Number(s) DE-NA0004144, 
the Department of Energy Office of Fusion Energy Sciences under Award Number DE-SC0022979, the University of Rochester, and the New York State Energy Research and Development Authority.
A.V.A. was supported by the U.S. National Science Foundation under Award No. PHY-2512067.
The support of DOE does not constitute an endorsement by DOE of the views expressed in this paper.
This research used the open-source particle-in-cell code WarpX \url{https://github.com/ECP-WarpX/WarpX}, primarily funded by the US DOE Exascale Computing Project. Primary WarpX contributors are with LBNL, LLNL, CEA-LIDYL, SLAC, DESY, CERN, and TAE Technologies. We acknowledge all WarpX contributors. This research used resources of the National Energy Research Scientific Computing Center, a DOE Office of Science User Facility supported by the Office of Science of the U.S. Department of Energy under Contract No. DE-AC02-05CH11231, using NERSC awards FES-ERCAP0028326 and FES-ERCAP0032999.
 
This report was prepared as an account of work sponsored by an agency of the U.S. Government. Neither the U.S. Government nor any agency thereof, nor any of their employees, makes any warranty, express or implied, or assumes any legal liability or responsibility for the accuracy, completeness, or usefulness of any information, apparatus, product, or process disclosed, or represents that its use would not infringe privately owned rights. Reference herein to any specific commercial product, process, or service by trade name, trademark, manufacturer, or otherwise does not necessarily constitute or imply its endorsement, recommendation, or favoring by the U.S. Government or any agency thereof. The views and opinions of authors expressed herein do not necessarily state or reflect those of the U.S. Government or any agency thereof.


\section*{Data Availability}

The data that support the findings of this study are available from the corresponding author upon reasonable request.

\section*{Author Contributions}

K.W. performed the numerical modeling and simulations. H.G.R. obtained funding. All the authors contributed to writing, editing, and conceptualization.

\section*{Competing interests}

The authors declare no competing interests.

\section*{References}

%


\end{document}